\documentclass[twocolumn]{aastex631}

\usepackage{amsmath,amssymb}
\usepackage{graphicx}
\usepackage{xfrac}

\usepackage{siunitx}
\usepackage[nameinlink]{cleveref}

\DeclareSIUnit{\mile}{mi} 

\graphicspath{{./}{figures/}}

\newcommand{\ssapy}{\textit{SSAPy}}

\shorttitle{Benchmark: One Million Cislunar Trajectories}
\shortauthors{Yeager et al.}

\begin{document}

\title{An Open Benchmark of One Million High-Fidelity Cislunar Trajectories}

\correspondingauthor{Travis Yeager}
\email{yeagerastro@gmail.com}

\author{Travis Yeager}
\affiliation{Lawrence Livermore National Laboratory, 7000 East Ave, Livermore, CA 94550, USA}

\author{Denvir Higgins}
\affiliation{Lawrence Livermore National Laboratory, 7000 East Ave, Livermore, CA 94550, USA}

\author{Peter McGill}
\affiliation{Lawrence Livermore National Laboratory, 7000 East Ave, Livermore, CA 94550, USA}

\author{Kerianne Pruett}
\affiliation{Lawrence Livermore National Laboratory, 7000 East Ave, Livermore, CA 94550, USA}

\author{Alexx Perloff}
\affiliation{Lawrence Livermore National Laboratory, 7000 East Ave, Livermore, CA 94550, USA}

\author{Tara Grice}
\affiliation{Lawrence Livermore National Laboratory, 7000 East Ave, Livermore, CA 94550, USA}

\author{Michael Schneider}
\affiliation{Lawrence Livermore National Laboratory, 7000 East Ave, Livermore, CA 94550, USA}

\begin{abstract}
Cislunar space spans from geosynchronous altitudes to beyond the Moon and will underpin future exploration, science, and security operations. We describe and release an open dataset of one million numerically propagated cislunar trajectories generated with the open-source Space Situational Awareness Python package (\ssapy). The model includes high-degree Earth/Moon gravity, solar gravity, and Earth/Sun radiation pressure; other planetary gravities are omitted by design for computational efficiency. Initial conditions uniformly sample commonly used osculating-element ranges, and each trajectory is propagated for up to six years under a single, fixed start epoch. The dataset is intended as a reusable benchmark for method development (e.g., space domain awareness, navigation, and machine-learning pipelines), a reference library for statistical studies of orbit families, and a starting point for community-driven extensions (e.g., alternative epochs). We report empirically observed stability trends (e.g., a band near $\sim$5 GEO and persistence of some co-orbital classes including L4/L5 librators) as dataset descriptors rather than new dynamical results. The chief contribution is the scale, fidelity, organization (CSV/HDF5 with full state time series and metadata), and open availability, which together lower the barrier to comparative and data-driven studies in the cislunar regime.
\end{abstract}


\section{Introduction}

Cislunar space, while increasingly significant in the context of space exploration and resource utilization, currently lacks a universally agreed-upon definition. A practical definition describes cislunar space as the region bounded by the altitude of Earth’s geosynchronous orbits on the lower end and the Earth--Moon system’s sphere of influence (SOI) on the upper end. The SOI of the Earth--Moon system can be approximated as
\begin{equation}
  r_{\text{SOI}}(\theta) \approx a \left( \frac{m}{M} \right)^{\frac{2}{5}} \frac{1}{\sqrt[10]{1 + 3\cos^2\!\theta}},
  \label{eq:sphere_of_influence}
\end{equation}
within which the Earth--Moon system exerts the dominant gravitational influence. Beyond the SOI, gravitational forces from other bodies become comparable, significantly affecting trajectories. The average radius of the Earth--Moon SOI is approximately \SI{929000}{\kilo\metre}, or about $2.4\times$ the average Earth--Moon distance.

Cislunar space has garnered increasing attention as governments and private organizations recognize its potential for exploration, resource utilization, and national security. When the United States Space Force (USSF) was introduced in 2019, it defined its operational ``sphere of influence'' as extending to \SI{272000}{\mile} (\SI{437742}{\kilo\metre}) and beyond---more than a tenfold increase in range and a thousand-fold expansion in service volume compared to traditional Earth-centric operations \citep{nasa2020spaceforce}. However, current missions and supporting infrastructure are largely confined to regions near the Moon, such as the L2 Lunar Lagrange point, located $\sim$\SI{66000}{\kilo\metre} beyond the Moon, closely aligning with the USSF operational sphere of interest.

The development of cislunar space offers numerous benefits. Future missions beyond the Earth--Moon system will rely on infrastructure established within cislunar space, making it a critical stepping stone for deep-space exploration. Cislunar space provides a unique low-gravity environment both accessible from Earth and sufficiently distant to enable a wide range of scientific, commercial, and strategic activities \citep{gerstenmaier2017progress,mozer2019future,butow2020state,holzinger2021primer}. Establishing a permanent human presence on the Moon will require efficient transportation, which depends on a robust understanding of cislunar orbital dynamics. These dynamics are complex, driven by the extended gravitational influence of the Moon, perturbations from the Sun and planets, and thermal radiation forces from Earth and the Sun, leading to trajectory deviations that compound to kilometer-level errors within days.

The operational complexity of cislunar space demands innovative approaches to mission planning and execution \citep{Whitley2016}. Recent literature highlights the need for robust Space Domain Awareness (SDA), navigation, and trajectory design frameworks to support an expanding ecosystem of crewed and robotic missions \citep{Whitley2016,Duffy2024}. Recent surveys articulate open questions for lunar science and cislunar infrastructure \citep{Lin2024,Cinelli2024} and explore observation and orbit-determination challenges for cislunar objects \citep{Hou2024}, underscoring the need for shared, high-fidelity datasets for method development and evaluation.

\noindent\textit{What this release provides.} A single-epoch, six-year, million-trajectory dataset with high-fidelity force modeling; complete state and observable time series; and standardized CSV/HDF5 formats to accelerate reproducible research in cislunar dynamics and SDA.

\subsection{The three-body problem}
The circular and elliptic restricted three-body problems (CR3BP/ER3BP) are widely used as simplified models for cislunar dynamics because they capture essential multi-body structure at modest computational cost. The CR3BP does not admit general closed-form (analytical) solutions; beyond equilibria and special symmetric cases, one relies on qualitative theory, numerical continuation, and direct integrations to compute families of periodic and quasi-periodic motions in the synodic frame \citep[e.g.,][]{holzinger2021primer}. When comparing such motions to ``Keplerian elements,'' we mean osculating elements with respect to a chosen primary (Earth or Moon); these elements vary in time and are generically quasi-periodic in an inertial frame because the fundamental frequencies are not, in general, commensurate with the synodic frequency. Consequently, CR3BP models are invaluable for insight and initial design, while high-fidelity, non-Keplerian propagation is required to assess long-term behavior under additional perturbations.

\section{\ssapy}
The orbital data used here were generated using \ssapy, a highly parallelizable and customizable Python package developed and open-sourced by LLNL \citep{ssapy,ssapyprep}. Built on C/C++ libraries with a Python wrapper, \ssapy{} simplifies high-fidelity orbital modeling and analysis for cislunar and low-Earth orbits. Users can define mass, area, and coefficients for drag and radiation, providing a robust framework for simulating dynamics in various environments.

To numerically integrate the equations of motion, \ssapy{} offers fixed-, variable-, and multi-step integrators so users can balance accuracy and cost for a given scenario. \ssapy{} supports LEO, GEO, and highly elliptical orbits, with selectable gravitational models from point-mass to high-degree harmonics. It incorporates atmospheric drag and radiation pressure with options ranging from basic approximations to advanced ray-tracing. Numerical integrators propagate initial conditions with controlled accuracy, and vectorized propagation improves performance and scalability. Accuracy has been validated against STK and GMAT. Documentation is hosted on the open-source repository.\footnote{\url{https://software.llnl.gov/SSAPy/}}%
\footnote{\url{https://github.com/LLNL/SSAPy}}

\section{Building the cislunar data}

The distance from Earth’s center to geosynchronous orbit is GEO. The average Earth--Moon distance is referred to as a Lunar Distance (LD), here \SI{384399}{\kilo\metre}.

A total of one million orbits were integrated, of which 54\% remained stable for at least one year and 9.7\% for six years. Orbits are deemed stable if they do not approach too close to the Earth or Moon and do not travel significantly far from the Earth--Moon system. Stability requires that an orbit never: (1) descends within the geosynchronous radius of Earth, (2) comes within two lunar radii of the Moon’s center, or (3) exceeds twice the Moon’s orbital distance from Earth. The lower limit of GEO and upper limit of $2$~LD span the range of most cislunar orbits. Beyond $2$~LD, orbital periods become significantly longer and are not useful for current needs. Excluding orbits that approach within two lunar radii of the Moon excludes lunar impacts.

\subsection{Initialization of the Cislunar Orbits}

Cislunar orbits are initialized using osculating Keplerian elements. There is no difference between initializing with $(\vec r,\vec v)$ or Keplerian elements; we use the latter due to common practice (e.g., TLEs). The six elements are semi-major axis $a$, eccentricity $e$, inclination $i$, true anomaly \textit{ta}, argument of periapsis \textit{pa}, and right ascension of the ascending node \textit{raan}. Vectors $\vec r$ and $\vec v$ are in GCRF.

Initial elements are uniformly sampled: $a\in[\text{GEO}=\SI{4.216e7}{m},\,2\text{LD}=\SI{7.688e8}{m}]$, $e\in[0,1)$, $i\in[0,\pi/2]$, and \textit{ta}, \textit{pa}, \textit{raan} $\in[0,2\pi)$.

The start date is 1980-01-01 00:00:00, setting the Solar System configuration (Earth, Moon, Sun). Because the Sun--Earth--Moon geometry is quasi-periodic, stability statistics do depend on epoch. We therefore treat this single-epoch dataset as a clearly defined slice of phase space for benchmarking and method development, rather than as an epoch-invariant census.

\paragraph{Epoch, frames, and constants.}
All initial conditions are defined at the single start epoch 1980-01-01 00:00:00 \emph{TT} in the GCRF/ICRF inertial frame. The Earth and Moon gravity fields are EGM2008 (degree/order 180) and GRGM1200A, respectively; the associated gravitational parameters are taken from those models as implemented in \ssapy. The Sun is modeled as a point mass with constants bundled with \ssapy (documented in the repository). Per-run attributes in the HDF5 files record the ephemeris/constants actually used, enabling exact reproducibility of state initialization and propagation settings.

\paragraph{Scope of the single-epoch baseline.}
We fix a single start epoch to create a controlled, reusable baseline for benchmarking algorithms and training models. Rather than claim epoch-invariant rates, we release the present dataset as a clearly defined baseline and provide initialization ranges, modeling settings, and file formats so the community can generate companion ensembles at additional epochs.

\subsection{Propagation Model}

Orbit propagation depends on satellite properties. Nominal \ssapy{} values: cross-sectional area $0.25~\mathrm{m^2}$, mass $250~\mathrm{kg}$, drag coefficient $2.3$, radiation pressure coefficient $1.3$ \citep{montenbruck2002satellite}. Area and mass affect radiation pressure and drag. Drag coefficient affects atmospheric drag only; radiation pressure coefficient applies to solar and terrestrial radiation pressure.

Net acceleration includes gravitational and thermal forces. The Earth uses EGM2008 (degree/order 180) \citep{pavlis2012development}; the Moon uses GRGM1200A \citep{Lemoine2014,Goossens2016}; the Sun is a point mass. Solar and Earth radiation pressure and atmospheric drag are included. Other planets’ gravity is excluded. High-fidelity modeling is warranted whenever the cumulative effect of neglected perturbations exceeds mission-relevant tolerances. In the cislunar regime, the dominant non-two-body accelerations (high-degree Earth/Moon gravity and radiation pressure) are $\mathcal{O}(10^{-5}\text{--}10^{-7})~\mathrm{m\,s^{-2}}$; over months to years these produce kilometer- to thousands-of-kilometers-scale deviations if omitted, which motivates our inclusion of those terms in the baseline propagations. Orbits are propagated with a fixed-step Runge--Kutta 7/8 and $\Delta t=\SI{10}{s}$. We record per-run tolerances and model parameters in the HDF5 attributes to make run-specific settings explicit in downstream use.

\subsection{Stopping Conditions}

\begin{figure}[t]
  \centering
  \includegraphics[width=\linewidth]{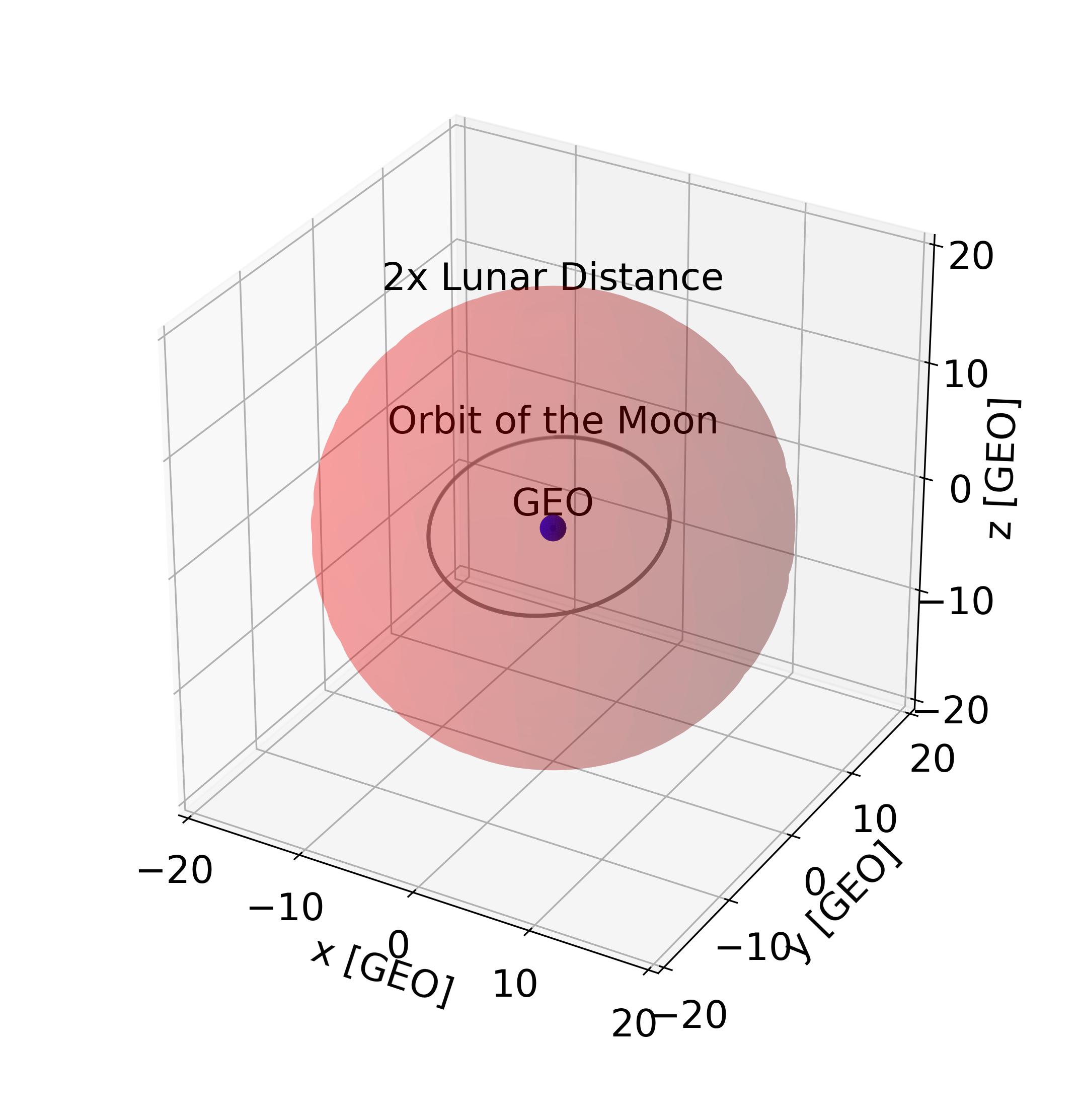}
  \caption{Cislunar volume to scale. The Moon’s orbit (gray), the geostationary shell (blue), and the outer bound at $2$~LD (red). Earth is centered at $(0,0)$ and not visible at this scale.}
  \label{fig:cislunar_volume}
\end{figure}

All orbits begin on 1980-01-01 00:00:00 and run up to six years. To continue propagation, an orbit must: (1) remain above GEO, (2) not collide with the Moon, and (3) not exceed $2$~LD from Earth. If violated, integration halts and the orbit is assigned a \emph{lifetime} (elapsed time from start). \Cref{fig:cislunar_volume} illustrates the allowed volume (red), the GEO shell (blue), and the Moon’s orbit (gray).

\subsection{Data Structure}

Two data structures organize the datasets. A CSV file contains one row per orbit with columns described in \Cref{tab:csv_columns}. An HDF5 file stores additional integration metadata and full time series; keys follow \texttt{orb\_id/key} with base keys in \Cref{tab:hdf5_keys}. A subset includes additional keys in \Cref{tab:hdf5_keys_additional}.

\begin{table*}[t]
\caption{CSV columns}
\label{tab:csv_columns}
\centering
\begin{tabular}{ll}
\hline
\textbf{Column} & \textbf{Description} \\
\hline
\texttt{orb\_id} & Orbital index to identify the orbit across all data \\
\texttt{ejection} & Reason for halting integration \\
\texttt{lifetime} & Duration of integration \\
\texttt{period} & Two-body (Earth-centered) Keplerian estimate from $a$ \\
\texttt{perigee} & Initial perigee \\
\texttt{apogee} & Initial apogee \\
\texttt{r0} & Initial position vector (GCRF) \\
\texttt{v0} & Initial velocity vector (GCRF) \\
\texttt{pm\_ra\_min/max} & Min/Max proper motion in right ascension \\
\texttt{pm\_dec\_min/max} & Min/Max proper motion in declination \\
\texttt{M\_v\_min/max} & Min/Max visual brightness \\
\texttt{a,e,i,tl,pa,raan,ta} & Initial Keplerian elements \\
\hline
\end{tabular}
\end{table*}

The \texttt{ejection} label can be \texttt{hit\_lowest\_alt}, \texttt{hit\_moon}, \texttt{leaves\_earth\_moon}, or \texttt{stable}. A flag \texttt{near\_moon} may be set if an otherwise stable orbit approaches the Moon within a distance equal to the instantaneous speed times one hour.

\begin{table*}[t]
\caption{HDF5 keys (base dataset)}
\label{tab:hdf5_keys}
\centering
\begin{tabular}{ll}
\hline
\textbf{Key} & \textbf{Description} \\
\hline
\texttt{orb\_id} & Orbit index \\
\texttt{ejection} & Reason for halting integration \\
\texttt{lifetime} & Duration of integration \\
\texttt{period} & Initial two-body Keplerian estimate \\
\texttt{perigee}, \texttt{apogee} & Initial perigee/apogee \\
\texttt{r}, \texttt{v} & Time series of position/velocity (GCRF) \\
\texttt{ra}, \texttt{dec}, \texttt{range} & Apparent RA/Dec and geocentric range vs.\ time \\
\texttt{pm\_ra}, \texttt{pm\_dec} & Proper motion vs.\ time \\
\texttt{M\_v} & Brightness vs.\ time \\
\texttt{semi\_major\_axis}, \texttt{eccentricity}, \texttt{inclination} & Osculating elements vs.\ time \\
\texttt{true\_longitude}, \texttt{argument\_of\_periapsis}, \texttt{longitude\_of\_ascending\_node} & Additional elements vs.\ time \\
\texttt{true\_anomaly} & Initial true anomaly \\
\texttt{r\_initial}, \texttt{v\_initial} & Initial GCRF state \\
\texttt{r\_earth\_min/max} & Closest/farthest geocentric distances \\
\texttt{vmin/vmax} & Min/Max speed \\
\texttt{r\_vmin/r\_vmax} & GCRF positions at \texttt{vmin/vmax} \\
\texttt{r\_moon\_min} & Minimum distance to the Moon \\
\hline
\end{tabular}
\end{table*}

For selected orbits, 25 ``nearby'' orbits are also integrated, initialized within \SI{10}{m} and \SI{1}{m\,s^{-1}} of the reference state, propagated for one month (or to the reference lifetime) with hourly outputs. The NumPy \texttt{cov} method (bias=\texttt{True}) yields covariance time series, from which standard deviation, median, mean, and maximum are computed.

\begin{table*}[t]
\caption{Additional HDF5 keys (select orbits)}
\label{tab:hdf5_keys_additional}
\centering
\begin{tabular}{ll}
\hline
\textbf{Key} & \textbf{Description} \\
\hline
\texttt{threebody\_nearby\_r/v} & Time series of position/velocity for 25 nearby orbits \\
\texttt{threebody\_covariances} & Covariance time series (nearby vs.\ reference) \\
\texttt{threebody\_std/median/mean/max\_divergence} & Statistics from covariance time series \\
\texttt{nearby\_r/v} & Position/velocity for 25 nearby orbits \\
\texttt{covariances} & Covariance time series \\
\texttt{std/median/mean/max\_divergence} & Statistics from covariance time series \\
\hline
\end{tabular}
\end{table*}

\section{Analysis of the cislunar data}

\begin{figure}[t]
  \centering
  \includegraphics[width=0.49\textwidth]{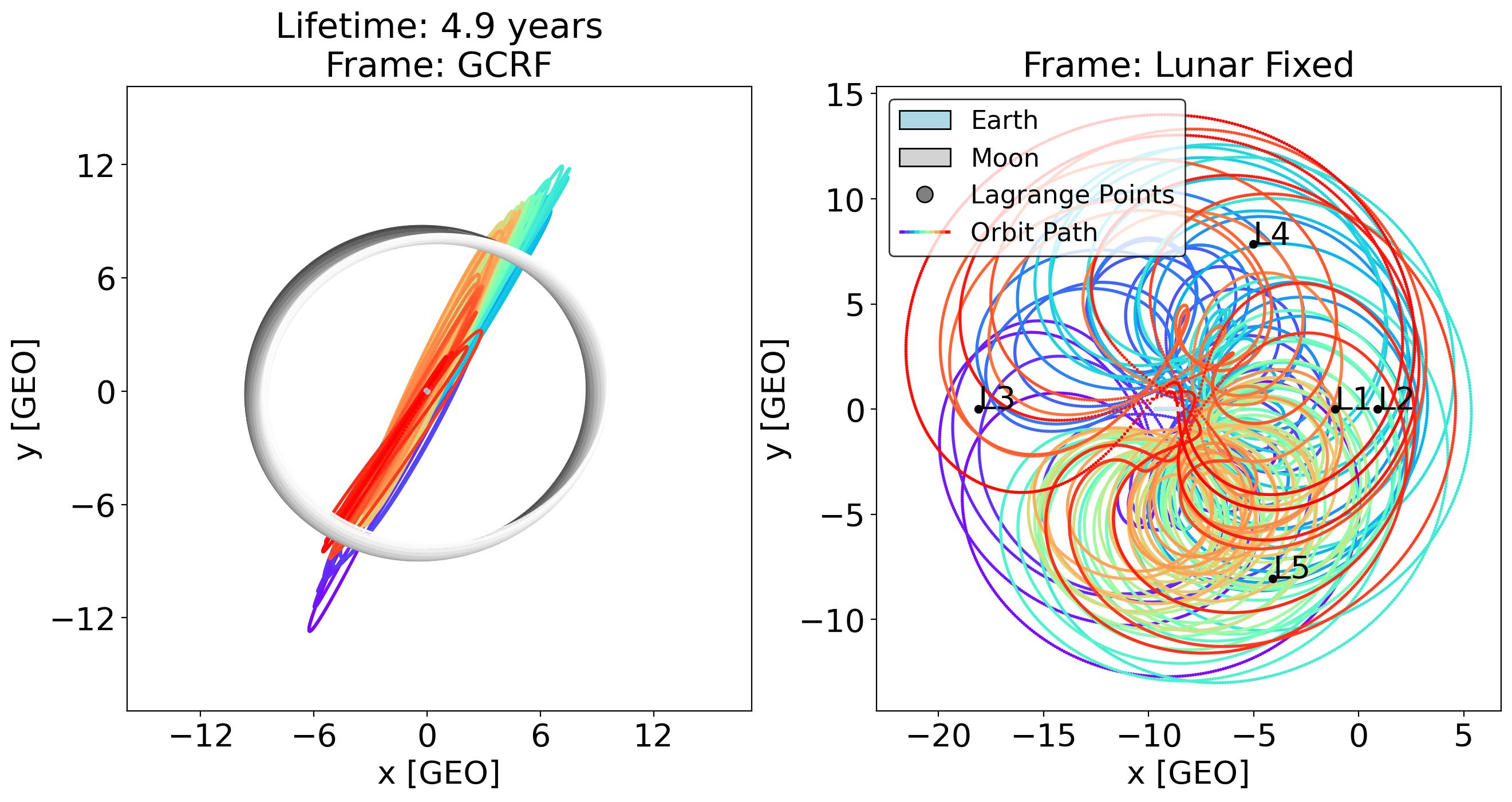}
  \caption{Example retrograde orbit (inclination $>90^\circ$) which drops below GEO before the six-year integration completes.}
  \label{fig:retrograde}
\end{figure}

\begin{figure}[t]
  \centering
  \includegraphics[width=0.49\textwidth]{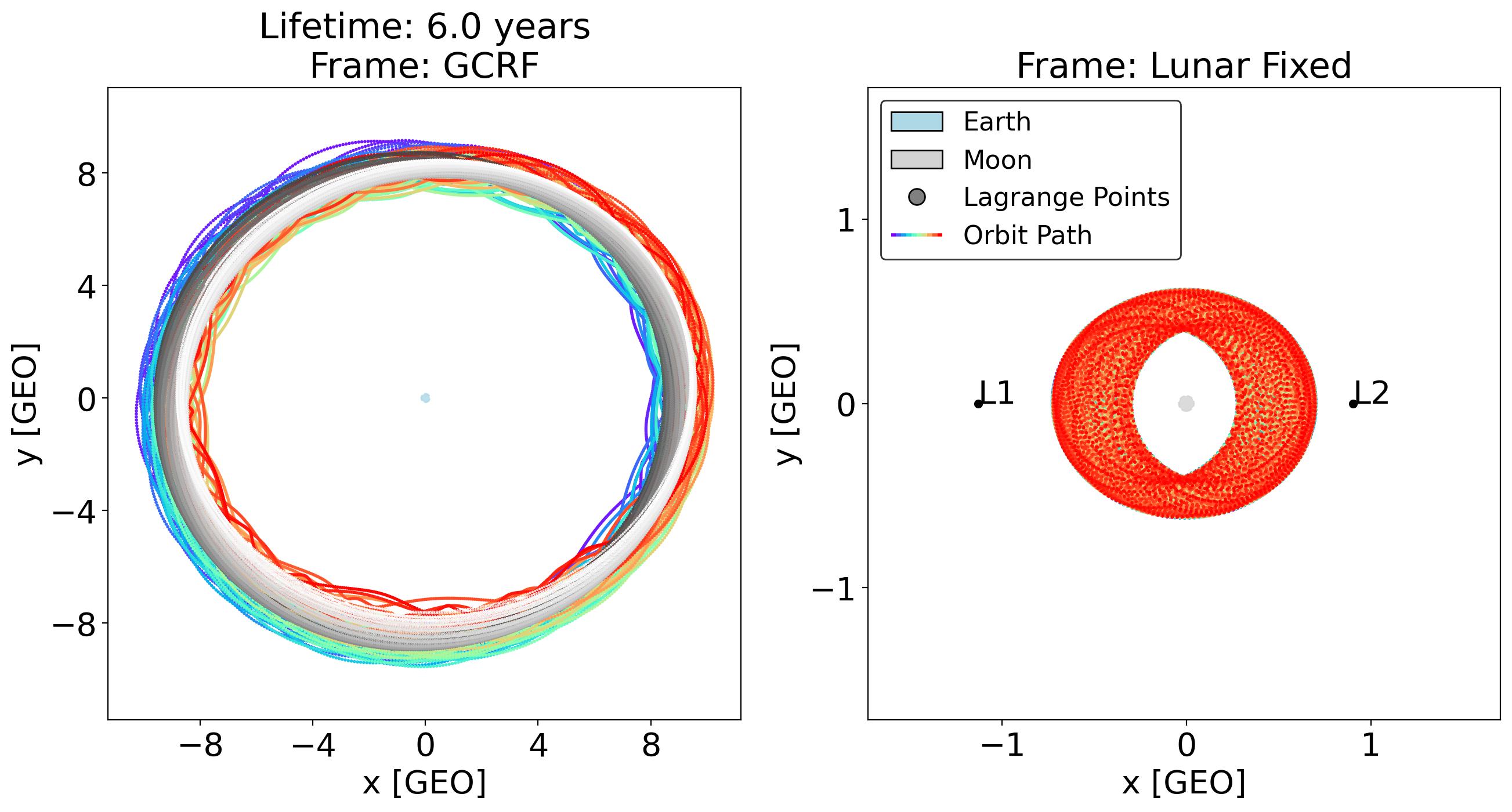}
  \caption{Example lunar orbit bound between the L1 and L2 Lagrange points.}
  \label{fig:mooncentric}
\end{figure}

\begin{figure}[t]
  \centering
  \includegraphics[width=0.49\textwidth]{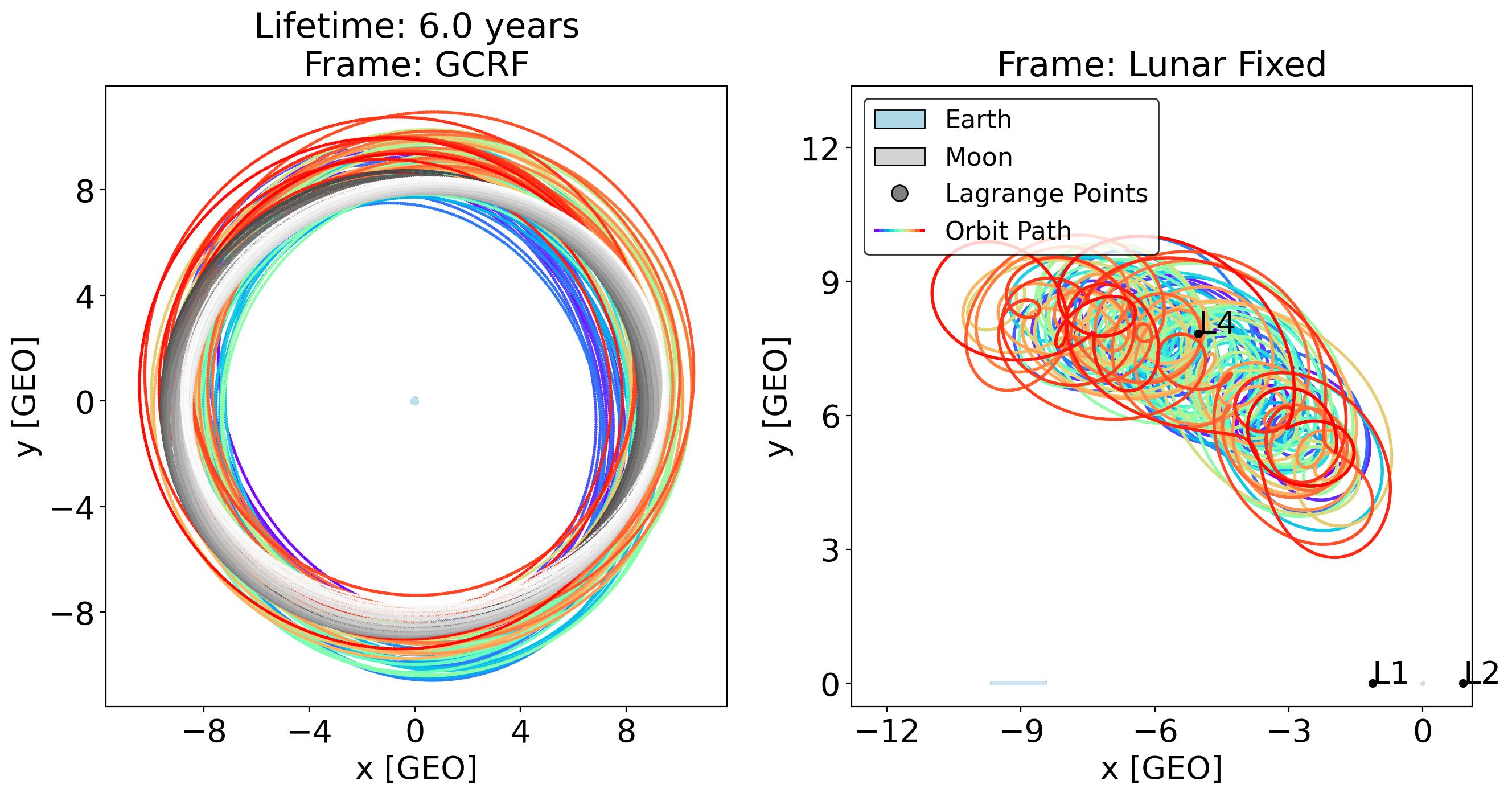}
  \caption{Example L4 orbit from the cislunar dataset.}
  \label{fig:l4_orbit}
\end{figure}

\begin{figure}[t]
  \centering
  \includegraphics[width=0.49\textwidth]{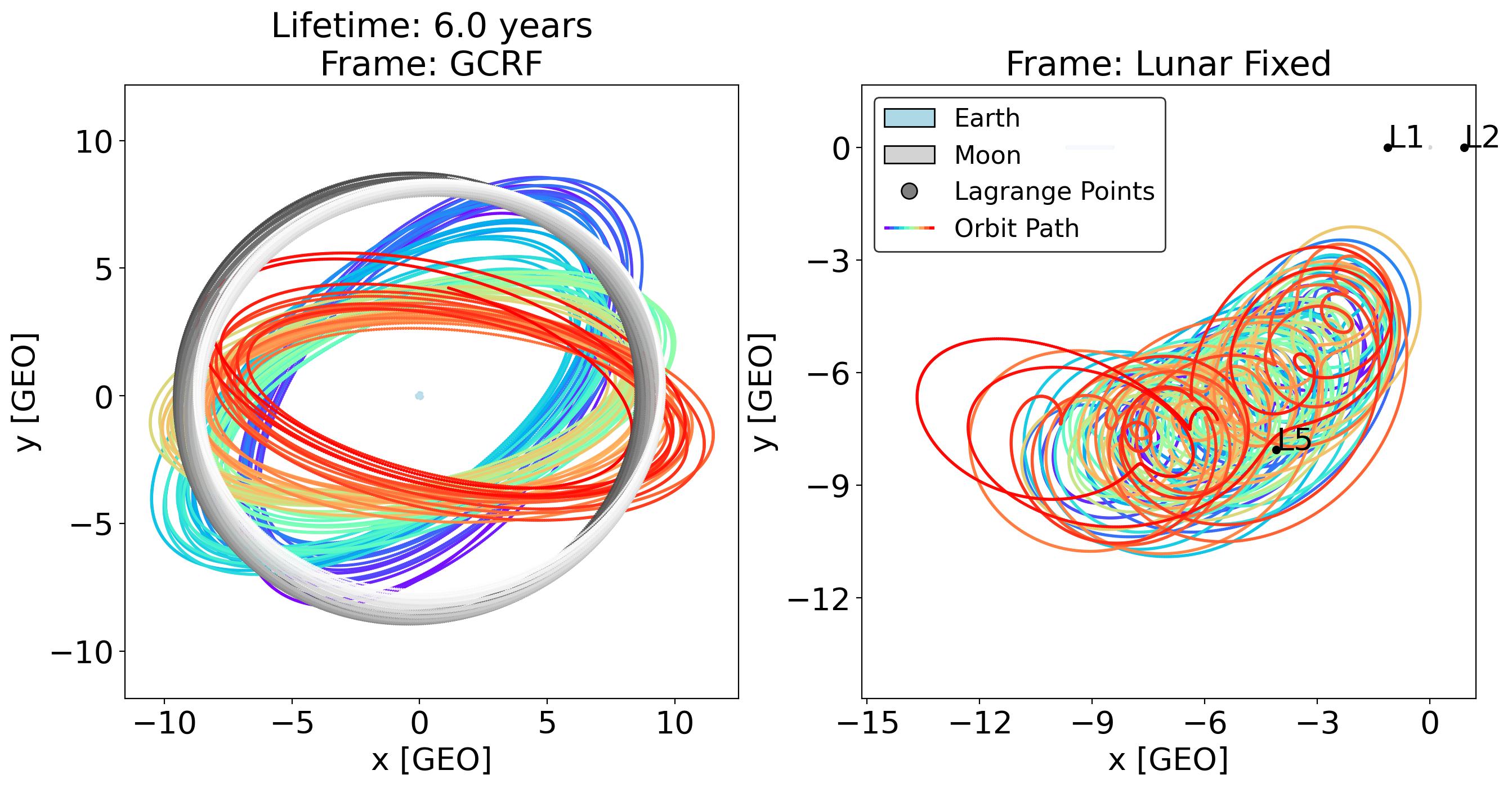}
  \caption{Example L5 orbit from the cislunar dataset.}
  \label{fig:l5_orbit}
\end{figure}

The examples in Figs.~\ref{fig:retrograde}--\ref{fig:l5_orbit} are intended as qualitative descriptors of the released dataset (not as claims of new dynamical families), illustrating orbit types commonly encountered in cislunar studies.

\begin{figure}[t]
  \centering
  \includegraphics[width=0.49\textwidth]{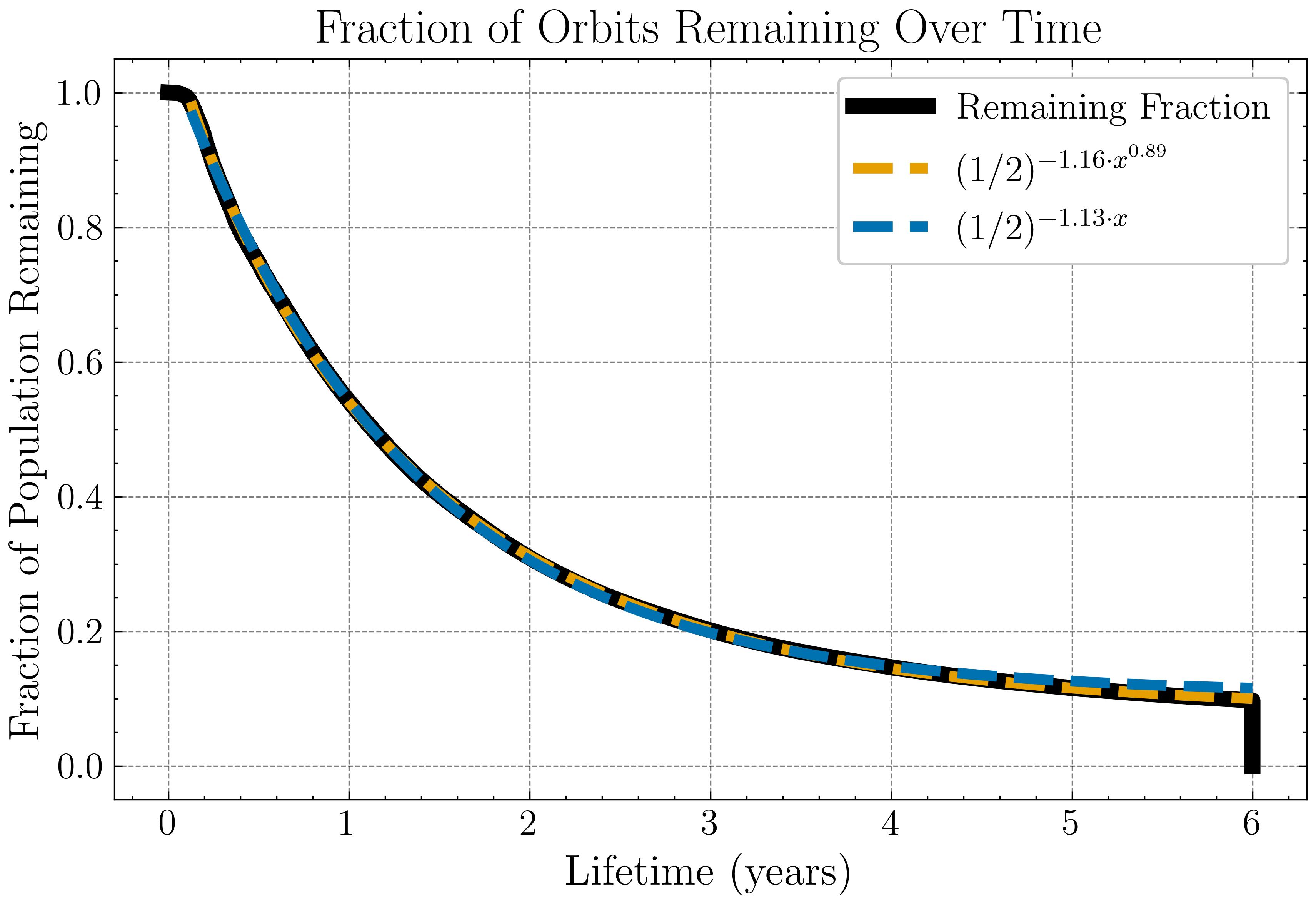}
  \caption{Fraction of the cislunar population remaining over time. Dashed: best-fit stretched half-life exponential; solid: half-life exponential.}
  \label{fig:lifetimes_hist}
\end{figure}

\Cref{fig:lifetimes_hist} shows the remaining fraction of the simulated population over six years. The best-fit trend lines (stretched exponential vs.\ pure exponential) are reported as dataset descriptors. The apparent band of longer lifetimes near $\sim$5 GEO and the persistence of some co-orbital classes (including librators near L4/L5) are consistent with prior qualitative expectations for multi-body structure; here we report them as characteristics of this particular single-epoch ensemble.

We write the simple exponential in half-life form,
\begin{equation}
  \frac{N(t)}{N_0} = \left(\frac{1}{2}\right)^{t/t_{1/2}} ,
  \label{eq:halflife}
\end{equation}
where $t_{1/2}$ is the time to half the population.

\begin{figure}[t]
  \centering
  \includegraphics[width=0.49\textwidth]{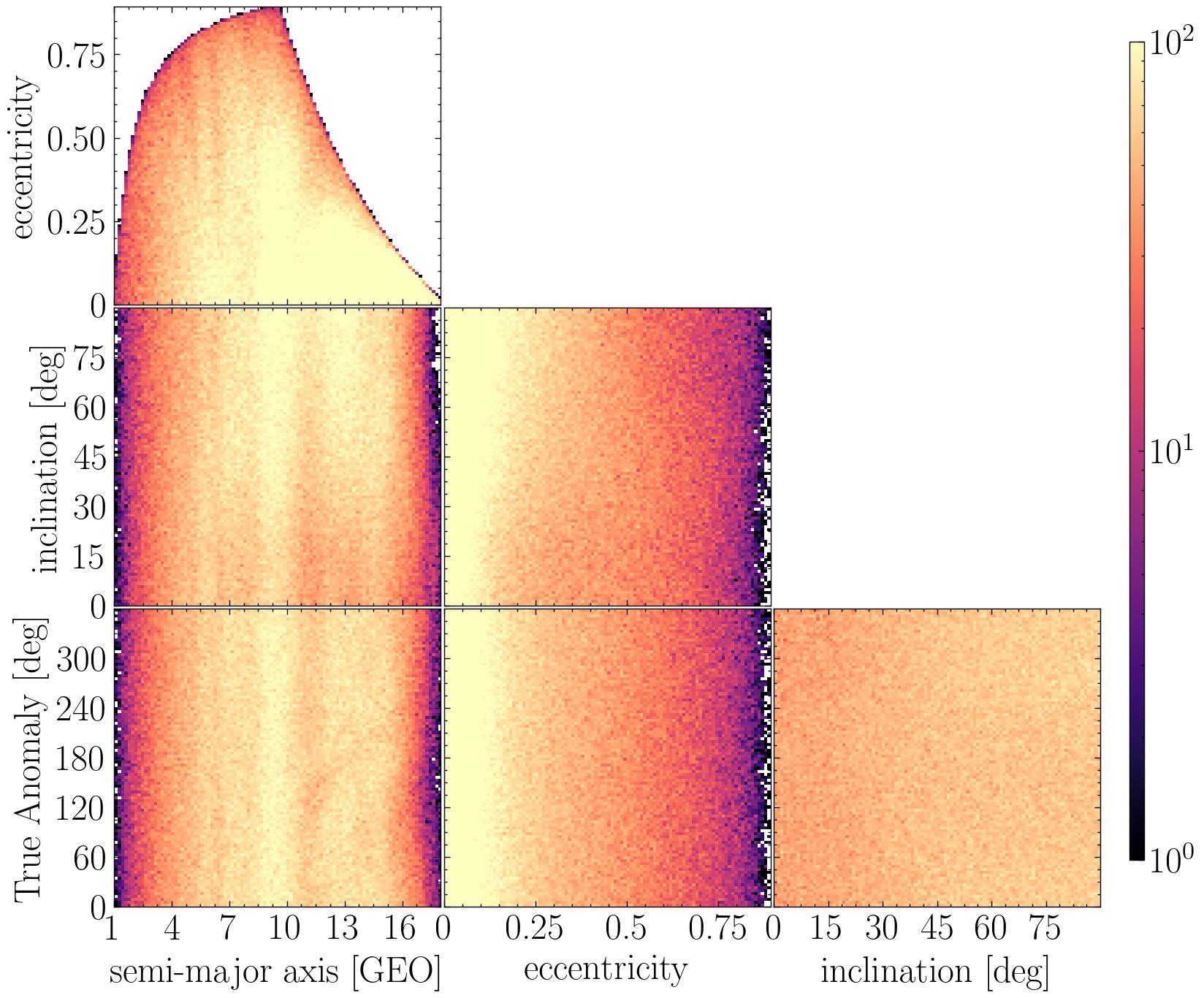}
  \caption{Initial osculating Keplerian elements for orbits remaining after one year.}
  \label{fig:1yr_stability_2dhist}
\end{figure}

\begin{figure}[t]
  \centering
  \includegraphics[width=0.49\textwidth]{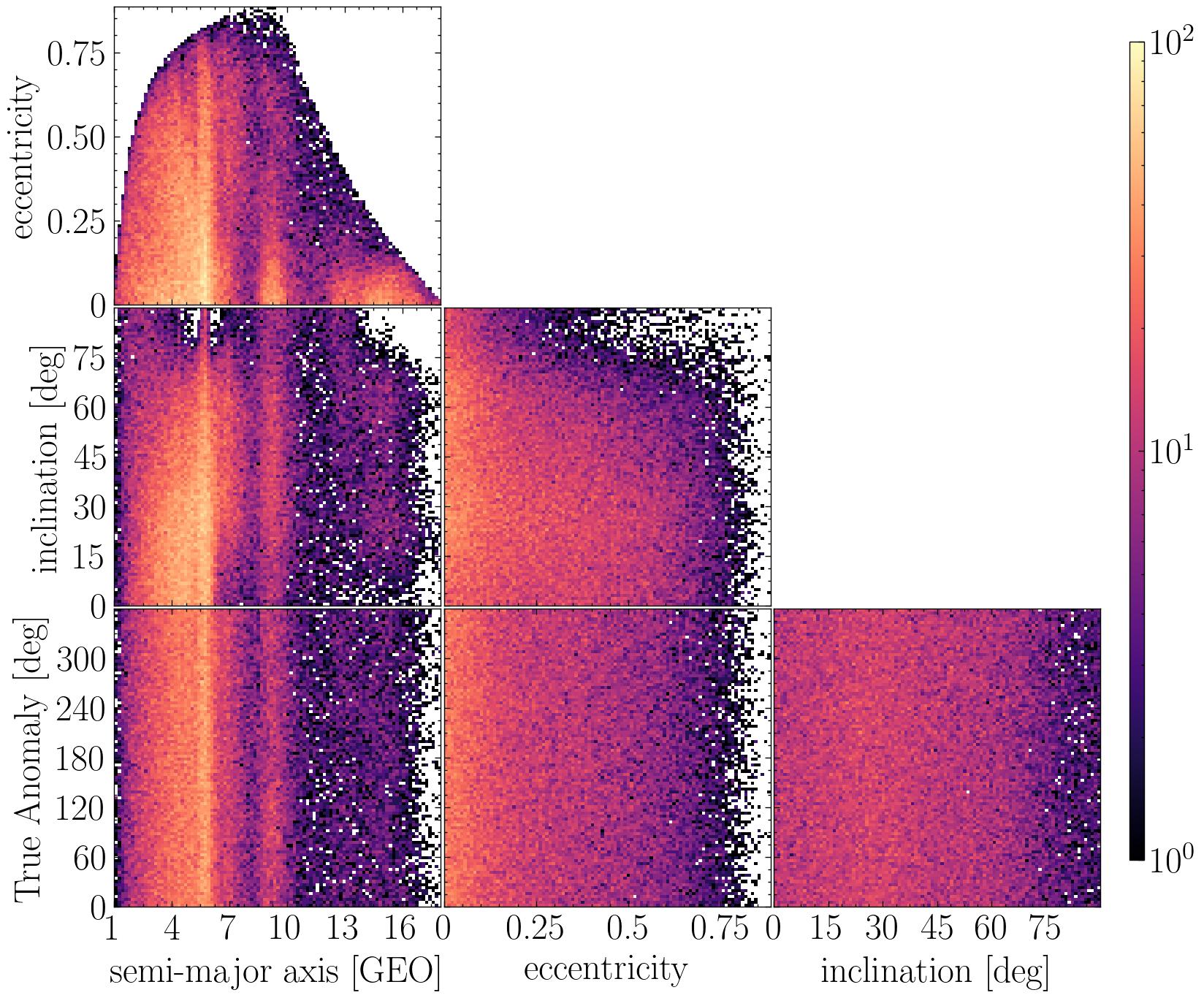}
  \caption{Initial osculating Keplerian elements for orbits remaining after six years.}
  \label{fig:6yr_stability_2dhist}
\end{figure}

\Cref{fig:1yr_stability_2dhist,fig:6yr_stability_2dhist} show 2D histograms of initial osculating elements for orbits stable to one and six years, respectively. After one year, a relationship between maximum eccentricity and semi-major axis emerges (\Cref{fig:1yr_stability_2dhist}, top). Lower $e$ is generally preferred, though stability with $e\lesssim0.25$ exists near the Moon’s orbit (9~GEO). Bands of reduced survival appear around 7--9 and 10--12~GEO, consistent with lunar perturbations. By six years (\Cref{fig:6yr_stability_2dhist}), these bands are pronounced; a line of stability persists near 9~GEO (lunar co-orbiters, including lunar-bound and Trojan orbits). A strong band of stability also appears near 5~GEO, just inside where the Moon’s gravity significantly perturbs orbits; this band admits higher stable eccentricities. Highly elliptical orbits require the least stationkeeping near $a\sim5$~GEO. Beyond the Moon, stable $e>0.25$ becomes sparse; at 13--16~GEO, only $e\lesssim0.15$ survive six years. The sharp upper-right boundary in $(a,e)$ is set by the $2$~LD outer limit.

\paragraph{Resonance context (qualitative).}
Bands of relative stability/instability in $(a,e,i)$ are plausibly linked to commensurabilities between the Earth-centered osculating mean motion and the Moon’s sidereal/synodic frequencies, as well as to lunar nodal/apsidal cycles. While we do not perform a resonance identification here, the semi-major-axis ranges exhibiting reduced survival are consistent with the expectation that near-resonant energy/inclination exchange with the Moon modulates lifetime. A detailed resonance map across epochs is a natural extension of this data paper.

\section{Accessing the cislunar data}

The dataset is hosted by LLNL on GDO: \href{https://gdo-cislunar.llnl.gov/}{https://gdo-cislunar.llnl.gov/}. The dataset will also be made available on the Unified Data Library; request the link from the corresponding author.

\section{Conclusion}

This paper’s primary contribution is an open, large-scale, and well-documented cislunar trajectory dataset designed to be a benchmark and training resource. Rather than claiming epoch-invariant stability rates or novel orbit families, we emphasize reproducibility, scale, and fidelity: high-degree Earth/Moon gravity, solar gravity, and radiation-pressure modeling; consistent initialization; and standardized CSV/HDF5 products with full state time series and metadata.

\paragraph{Future extensions.}
Two natural extensions, orthogonal to this data release, are (i) multi-epoch ensembles that span key Sun--Earth--Moon geometries (e.g., new/full moon and quadratures across seasons) and (ii) complementary long-horizon, lower-fidelity runs that probe asymptotic statistics. We view the present release as a controlled baseline to which such ensembles can be compared.

\nolinenumbers

\section{Acknowledgements}

This work was performed under the auspices of the U.S. Department of Energy by Lawrence Livermore National Laboratory under Contract DE-AC52-07NA27344 and was supported by the LLNL-LDRD Program under Project 22-ERD-054. Computing support came from the LLNL Institutional Computing Grand Challenge program.

\bibliographystyle{aasjournal}
\bibliography{refs}

\end{document}